\documentclass[epj]{svjour}
\usepackage{graphics,amsmath,amssymb,multirow,color,pstricks,pst-plot}
\usepackage{xspace,graphicx,comment,hyperref,cite}

\newcommand{\QQbar}{\ensuremath{Q\overline{Q}}\xspace}
\newcommand{\pt}{\ensuremath{p_{\rm T}}\xspace}
\newcommand{\dd}{\ensuremath{{\rm d}}\xspace}
\newcommand{\jpsi}{\ensuremath{{\rm J}/\psi}\xspace}
\newcommand{\psip}{\ensuremath{\psi{\rm (2S)}}\xspace}

\newcommand{\pTovM}{\ensuremath{p_{\rm T}/M}\xspace}
\newcommand{\lth}{\ensuremath{\lambda_\vartheta}\xspace}

\newcommand{\abscosth}{\ensuremath{|\cos\vartheta|}\xspace}

\newcommand{\Chic}{\ensuremath{\chi_{c}}\xspace}
\newcommand{\Chib}{\ensuremath{\chi_{b}}\xspace}
\newcommand{\chic}[1]{\ensuremath{\chi_{{c}#1}}\xspace}
\newcommand{\ChicZero}{\ensuremath{\chic{0}}\xspace}
\newcommand{\ChicOne}{\ensuremath{\chic{1}}\xspace}
\newcommand{\ChicTwo}{\ensuremath{\chic{2}}\xspace}
\newcommand{\ChicJ}{\ensuremath{\chic{J}}\xspace}
\newcommand{\lthOne}{\ensuremath{\lambda_\vartheta^{\chic{1}}}\xspace}
\newcommand{\lthTwo}{\ensuremath{\lambda_\vartheta^{\chic{2}}}\xspace}
\newcommand{\ROne}{\ensuremath{R^{\chic{1}}}\xspace}
\newcommand{\RTwo}{\ensuremath{R^{\chic{2}}}\xspace}
\newcommand{\lthpsi}{\ensuremath{\lambda_\vartheta^{\jpsi}}\xspace}

\newlength{\digitwidth} 

\begin{document}

\title{From prompt to direct \jpsi production:\\
new insights on the \ChicOne and \ChicTwo polarizations and feed-down contributions 
from a global-fit analysis of mid-rapidity LHC data}

\author{Pietro Faccioli\inst{1} 
\and Carlos Louren\c{c}o\inst{2}
\and Thomas Madlener\inst{3}
}
\institute{LIP, Lisbon, Portugal
\and 
CERN, Geneva, Switzerland
\and 
DESY, Hamburg, Germany}

\date{Received: May 12, 2020 / Revised version: date}
\abstract{
While the prompt \jpsi cross section and polarization have been 
measured with good precision as a function of transverse momentum, \pt, 
those of the directly produced \jpsi are practically unknown, given that the
cross sections and polarizations of the \ChicOne and \ChicTwo mesons,
large indirect contributors to \jpsi production, are only known with rather poor accuracy.
The lack of precise measurements of the \ChicJ polarizations induces 
large uncertainties in the level of their feed-down contributions to the prompt \jpsi yield, 
because of the polarization-dependent acceptance corrections.
The experimental panorama of charmonium production can be significantly 
improved through a consistent and model-independent global analysis 
of existing measurements of \jpsi, \psip and \Chic cross sections and 
polarizations, faithfully respecting all the correlations and uncertainties.
In particular, it is seen that the \ChicJ polarizations and feed-down 
fractions to \jpsi production have a negligible dependence on the \jpsi \pt, 
with average values
$\lthOne = 0.55 \pm 0.23$,  $\lthTwo = -0.39 \pm 0.22$, 
$\ROne = (18.8 \pm 1.4)\%$ and $\RTwo = (6.5 \pm 0.5)\%$.
The analysis also shows that $(67.2 \pm 1.9)$\% of the prompt \jpsi yield
is due to directly-produced mesons, of polarization constrained to
remarkably small values, $\lthpsi = 0.04 \pm 0.06$.
}

\titlerunning{\ChicOne and \ChicTwo polarizations and feed-down contributions to \jpsi production}

\maketitle

\section{Introduction}

The study of the production of heavy quarkonia,
bound states of charm or beauty quarks and antiquarks,
offers the best experimental laboratory to understand how quarks combine into hadrons, 
the least understood sector of quantum chromodynamics (QCD), 
the theory of strong interactions~\cite{bib:NRQCD,bib:QWGYR,bib:QWG2011}.
Indeed, the heaviness of these quarks implies that they have relatively small velocities in the \QQbar system,
so that their production can be theoretically described in two steps:
a short-distance regime where the \QQbar is produced (typically through gluon fusion), 
calculable using perturbative QCD,
followed by an intrinsically non-perturbative (long-distance) transition,
particularly challenging to understand at the present moment.

While it is known since long that experimental measurements of quarkonium cross sections and polarizations
should, in principle, lead to significant progress in our QCD-based understanding of hadron formation,
that prospect has faced serious hurdles for a long time, 
first because of several challenges in the execution of the measurements 
and the poor reliability of the resulting data~\cite{bib:EPJC69},
and second because most theoretical analyses of the data have not properly taken into consideration 
the correlations and uncertainties affecting the measurements,
as explained in Refs.~\cite{bib:FaccioliPLB736,bib:Faccioli:PLB773}.

The high-quality measurements made over the last decade at the LHC,
with a remarkable level of detail and precision, 
provided a much improved experimental situation.
In particular, double-differential cross sections, 
in transverse momentum, \pt, and rapidity, $y$, 
have been measured in pp collisions at $\sqrt{s} = 7$, 8 and 13\,TeV
for the \jpsi, \psip and $\Upsilon$(nS) vector states, by
ATLAS~\cite{bib:ATLASdimuon,bib:ATLASpsi2S,bib:ATLASYnS},
CMS~\cite{bib:BPH-14-001,bib:BPH-12-006,bib:BPH-15-005}
and LHCb~\cite{bib:LHCb_psi_cs,bib:LHCb_psip_cs,bib:LHCb_Upsilon_7_8,
bib:LHCb_psi_Upsilon_8,bib:LHCb_jpsi_13,bib:LHCb_psip_7_13,bib:LHCb_Upsilon_13}.
Also the polarizations have been measured for these states,
at mid-rapidity by CMS~\cite{bib:BPH-13-003,bib:BPH-11-023}
and at forward rapidity by 
LHCb~\cite{bib:LHCb_psi_pol,bib:LHCb_psip_pol,bib:LHCb_Upsilon_pol}.
In comparison with the very significant experimental progress made at the LHC 
regarding the differential cross sections and polarizations of vector quarkonia,
the corresponding knowledge of the \Chic and \Chib states has remained rather poor,
limited until recently to cross sections or cross section ratios affected by relatively large 
uncertainties~\cite{bib:ATLASchic,bib:BPH-11-010,bib:BPH-13-005,bib:LHCb_chic12_7,bib:LHCb_chic_7}.
First experimental measurements on the polarizations of the \ChicOne and \ChicTwo states 
have recently been reported by CMS~\cite{bib:BPH-13-001}, 
effectively constraining the \emph{difference} between the polarizations of the two states
but leaving their individual values mostly unknown.

The polarization of the produced particle is not only, by itself, 
a crucial element for the understanding of the underlying formation mechanisms, 
but is also an essential element for the determination of the acceptance correction 
of the corresponding cross section measurement. 
In fact, all the LHC collaborations published their cross section 
measurements~\cite{bib:ATLASdimuon,bib:ATLASpsi2S,bib:ATLASYnS,
bib:BPH-14-001,bib:BPH-12-006,bib:BPH-15-005,
bib:LHCb_psi_cs,bib:LHCb_psip_cs,bib:LHCb_Upsilon_7_8,
bib:LHCb_psi_Upsilon_8,bib:LHCb_jpsi_13,bib:LHCb_psip_7_13,bib:LHCb_Upsilon_13,
bib:ATLASchic,bib:BPH-11-010,bib:BPH-13-005,bib:LHCb_chic12_7,bib:LHCb_chic_7}
together with tables providing correction factors that reflect 
how the central values of each measurement change when the detection acceptance
(computed, by default, for unpolarized production)
is recomputed for other (extreme) polarization scenarios. 
One cannot underestimate the crucial importance of this knowledge, 
as the corrected production yields can vary by more than 50\% 
depending on the polarizations assumed in the evaluation of the acceptances,
an effect that vastly dominates over the statistical and systematic measurement uncertainties.
A consequence of the incomplete experimental information on the \ChicOne and \ChicTwo polarizations is,
therefore, that also their cross sections (and cross section ratios) continue to carry a large associated uncertainty.

This lack of knowledge also has a strong effect on the understanding of \jpsi production.
In fact, while the \emph{prompt} \jpsi differential cross section and polarization 
are the most precisely measured observables among all measurements in the field of quarkonium production, 
their interpretation in terms of properties of the physically-relevant \emph{directly} produced \jpsi mesons 
remains obscured by a large uncertainty, 
given the very significant and not well known fraction of \emph{indirect} production 
from \Chic feed-down decays.

This uncertainty also blurs the pattern of how the production of quarkonia of different masses, 
binding energies and quantum numbers is modified by the QCD medium produced in 
high-energy heavy-ion collisions~\cite{bib:Faccioli:Fate}.
Indeed, the feed-down fractions from heavier states are a crucial ingredient 
in the observation of signatures of the sequential suppression mechanism,
according to which the production rate of quarkonium states should be progressively suppressed, 
as the temperature of the medium increases,
following a hierarchy in the binding energy of the state~\cite{bib:DPS, bib:KKS}.

In this paper we report the results of a global fit of 
mid-rapidity charmonium measurements made by ATLAS and CMS at 7\,TeV, 
including the recent \Chic polarization measurement, 
to derive the best possible determinations of 
the \Chic-to-\jpsi feed-down fractions and \Chic polarizations, 
and also of the properties of direct \jpsi production. 
The analysis is completely independent of any quarkonium production theoretical model. 
It only relies on the published measurements, 
which include the indirect constraints that the differences between the \jpsi and \psip data 
impose on the \Chic cross sections and polarizations, 
through the feed-down contributions present in the \jpsi case and absent in the \psip case.
The transition probabilities from heavier to lighter states needed in this work 
are all well known and listed in the PDG tables~\cite{bib:PDG}. 
Our results are, therefore, fully data driven.

Besides reporting the \Chic polarizations and feed-down contributions to \jpsi production
determined by the global fit, 
we also discuss how improved constraints on the \Chic polarizations 
could be obtained with new charmonium measurements 
(other than direct measurements of the \Chic polarizations themselves).

\section{Experimental data and fit parametrization}

\begin{figure}[h]
\centering
\includegraphics[width=0.98\linewidth]{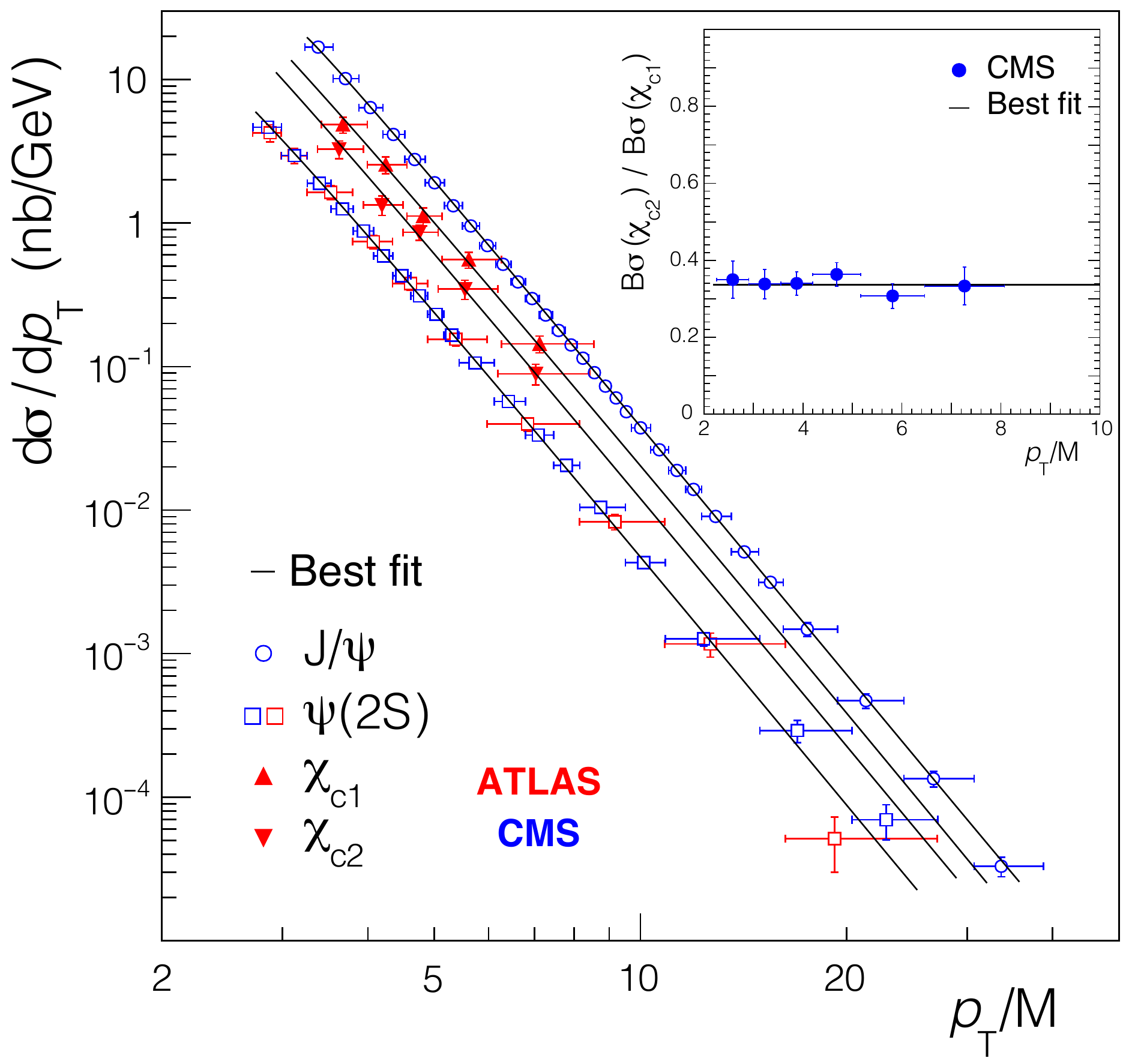}
\caption{Mid-rapidity prompt quarkonium cross sections measured in pp collisions
at $\sqrt{s} = 7$\,TeV
by ATLAS (red markers)~\cite{bib:ATLASpsi2S, bib:ATLASchic}
and CMS (blue markers)~\cite{bib:BPH-14-001}.
The inset shows the \ChicTwo to \ChicOne cross section ratio~\cite{bib:BPH-11-010}.
The curves represent the result of the fit described in the text.}
\label{fig:xsections}
\end{figure}

\begin{figure}[t!]
\centering
\includegraphics[width=0.94\linewidth]{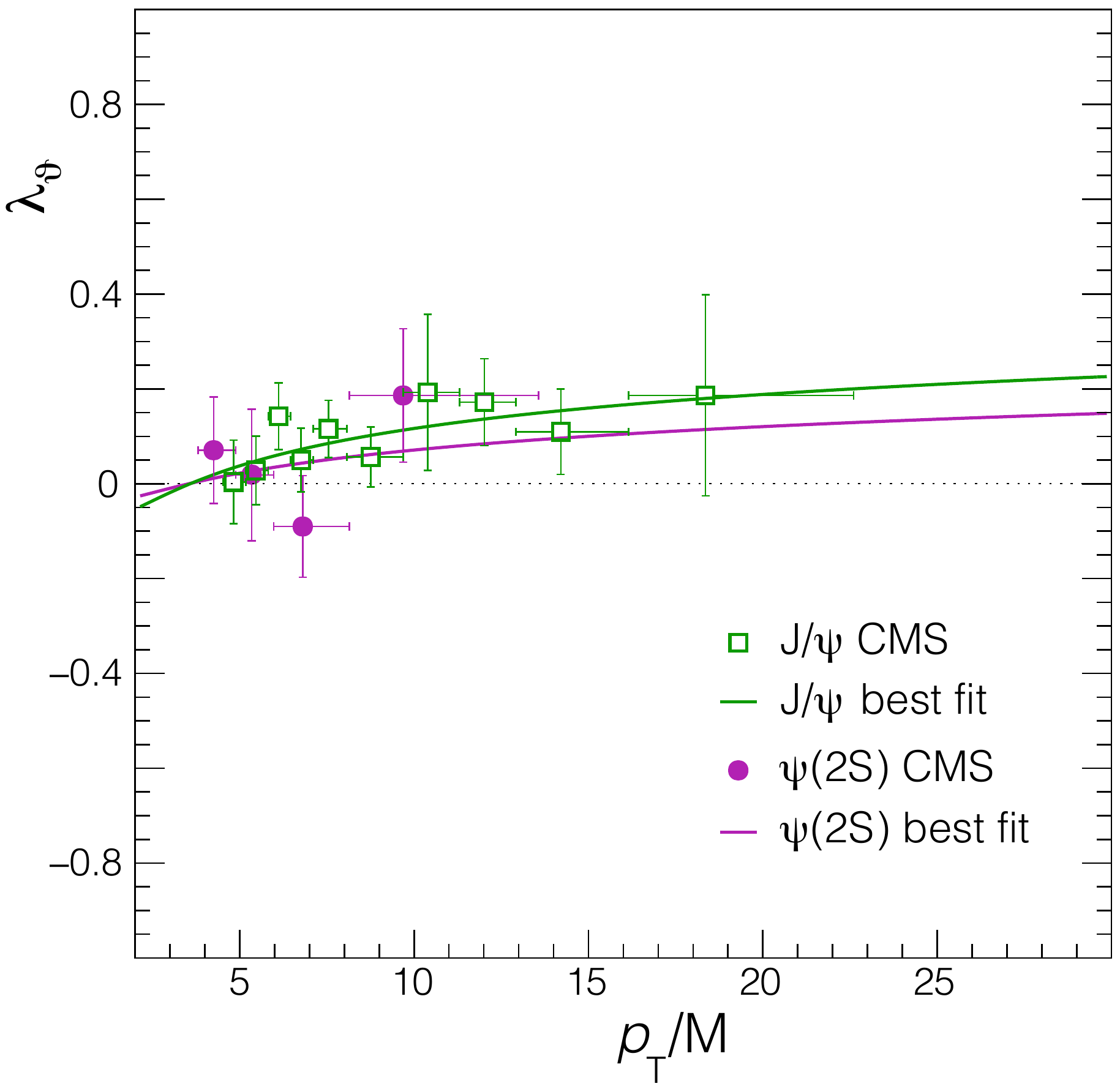}
\caption{Polar anisotropy parameter \lth, in the helicity frame,
measured by CMS in pp collisions at $\sqrt{s} = 7$\,TeV,
for prompt \jpsi and \psip dimuon decays~\cite{bib:BPH-13-003}.
Values corresponding to two (\jpsi) or three (\psip) rapidity bins were averaged.
The curves represent the result of the fit described in the text.}
\label{fig:lth}
\vglue15pt
\centering
\includegraphics[width=0.93\linewidth]{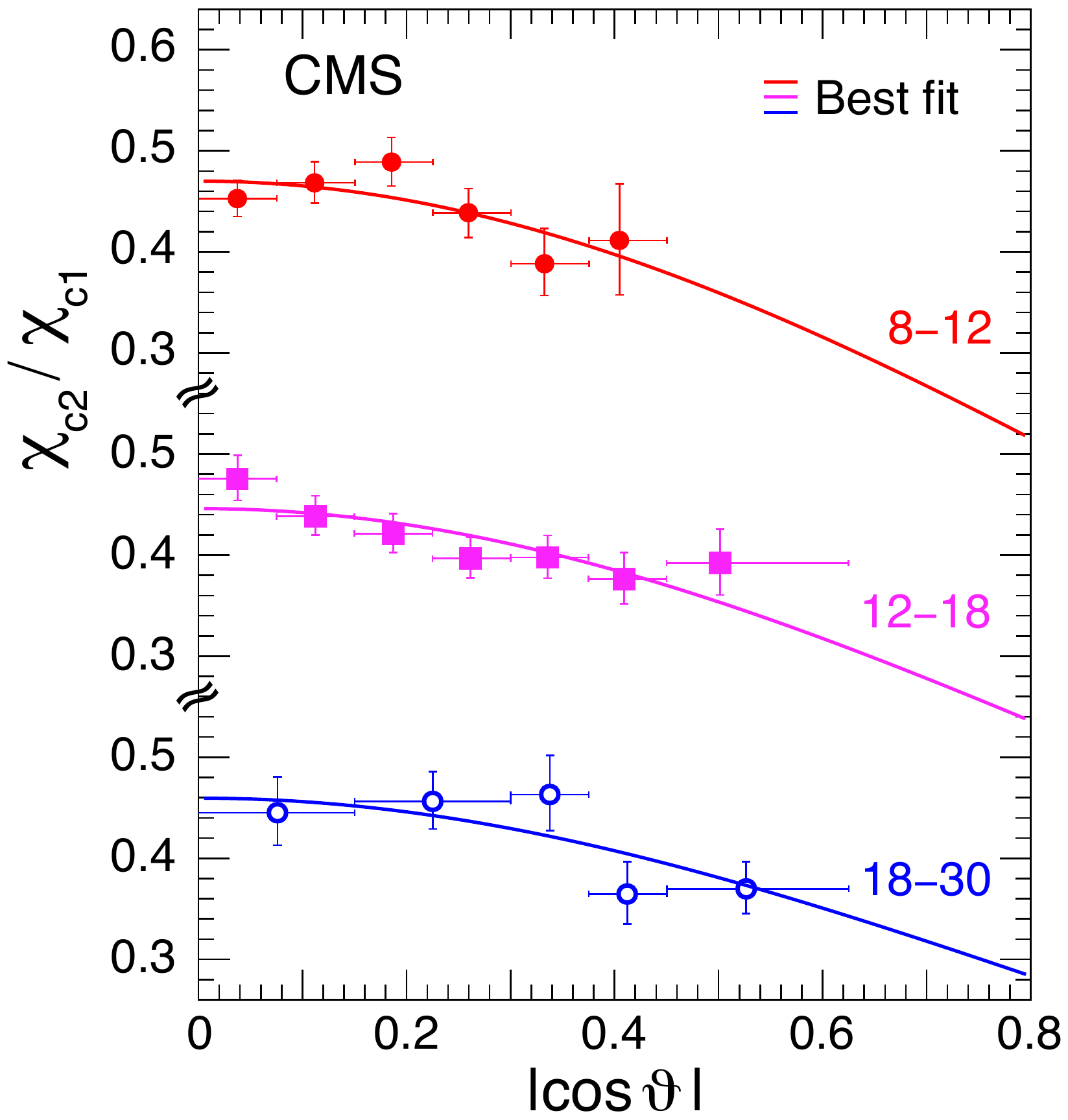}
\caption{The $\ChicTwo / \ChicOne$ yield ratio vs.\ \abscosth (in the helicity frame), 
for three \jpsi \pt bins (8--12, 12--18 and 18--30\,GeV),
as measured by CMS in pp collisions at $\sqrt{s} = 8$\,TeV~\cite{bib:BPH-13-001}.
The curves represent the result of the fit described in the text.}
\label{fig:Rchic_vs_costh}
\end{figure}

The data considered in our analysis are the \jpsi, \psip, \ChicOne and \ChicTwo 
differential cross sections measured by ATLAS and CMS 
at 7\,TeV~\cite{bib:ATLASpsi2S,bib:BPH-14-001,bib:ATLASchic}, 
as well as the \ChicTwo over \ChicOne cross section ratio~\cite{bib:BPH-11-010}
and the \jpsi and \psip polarizations~\cite{bib:BPH-13-003}
measured by CMS at the same energy.
All of these measurements have been reported as functions of \pt. 
To constrain the \ChicJ polarizations we also include the recent CMS measurement 
of the \ChicTwo over \ChicOne yield ratio versus \abscosth
($\vartheta$ being the lepton emission angle in the rest frame of the daughter \jpsi) 
in three \jpsi \pt bins.
The total number of independent data points is 108 and they are all
shown in Figs.~\ref{fig:xsections}, \ref{fig:lth} and~\ref{fig:Rchic_vs_costh}.

The \jpsi and \psip \pt-differential cross sections measured by ATLAS 
in the dimuon decay channel~\cite{bib:ATLASdimuon}
have not been included in our global-fit analysis because the data, 
reported in eight equidistant $|y|$ bins in the range $|y| < 2$,
show shapes as a function of \pt that vary quite strongly among the $|y|$ bins.
These variations, in particular between the $|y| < 0.25$ and $0.25 < |y| < 0.5$ bins,
are clearly not statistical fluctuations and
significantly exceed what one could expect from the reported (systematic) uncertainties,
pointing to an internal inconsistency affecting these two data sets.

To derive global observables from the analysis, 
a parametrization of yields and polarizations is necessary, 
because the kinematic binning of the reported distributions is not identical among all data sets and quarkonium states. 
Moreover, for the comparison/combination of results concerning objects of different mass scales 
the absolute transverse momentum, in which the data are binned, is not the best variable: 
it is, in fact, preferable to use a relative, dimensionless variable, in our case chosen as \pTovM, 
the ratio between the \pt and the mass of the quarkonium state. 
The convenience of this variable will become apparent in the next paragraphs, 
where we sometimes use the definition $\xi \equiv \pTovM$ to simplify the notation.

The \pTovM dependences of the considered particle yields are parametrized 
using a shape function $g(\xi)$,
normalized to unity at the arbitrary reference point $\xi^{*} = 5$,
a value close to the centre of gravity of the data:
$g(\xi) = h(\xi) / h(\xi^{*})$, with
\begin{equation}
\label{eq:powerLaw}
h(\xi) = \xi \cdot \bigg( 1+\frac{1}{\beta-2} \cdot \frac{\xi^2}{\gamma} \bigg)^{-\beta}\, .
\end{equation}
This functional shape describes very well the quarkonium transverse momentum distributions 
in different kinematic domains~\cite{bib:HERAb,bib:Faccioli:PLB773,bib:Faccioli:EPJC78p118}.
The parameter $\gamma$ (having the meaning of the average \pTovM\ squared) 
defines the function in the low-\pt turn-on region and is only mildly sensitive to the data we are considering; 
hence, only one $\gamma$ parameter will be considered, 
common for all states and polarization configurations and treated as global free parameter of the fit.
The power-law exponent $\beta$ describes the asymptotic high-\pt behaviour:
$g \propto \xi^{-(2\beta-1)}$ for $\xi \gg \sqrt{\gamma\, (\beta-2)}$.
It is, therefore, the shape parameter actually characterising each considered (sub-)process.

To ensure correct feed-down relations between the charmonium family members, 
we include a detailed account of how the mother's momentum and polarization 
are transferred to the daughter in the relevant decays:
$\psip \to \chi_{c1,2} \; \gamma$;
$\psip \to \jpsi \; X$;
$\chi_{c1,2} \to \jpsi \; \gamma$.
The rule for the momentum propagation from mother to daughter is, on average,
$\pt/m = P_{\rm T}/M$, where $M$ ($m$) and $P_{\rm T}$ (\pt) are, respectively,
the mass and laboratory transverse momentum of the mother (daughter) particle~\cite{bib:Faccioli:PLB773}.
The polarization transfer rules were calculated in the electric dipole approximation and
precisely account for the observable dilepton distribution with no need of higher-order terms~\cite{bib:chiPol}.
In particular, the \lthOne and \lthTwo (polar anisotropy) parameters 
refer to the shapes of the corresponding daughter-\jpsi's dilepton decay distributions,
which are the ones directly measured and fully reflect the \Chic polarization state,
while being insensitive to the uncertain contributions of higher-order photon multipoles~\cite{bib:chiPol}. 
Consequently, the terms \emph{longitudinal} and \emph{transverse} will here always refer to 
the yields of events where the daughter \jpsi has, respectively, angular momentum projection 
$J_z = 0$ ($\lth = -1$) and $J_z = \pm 1$ ($\lth = +1$), 
even if the mother \ChicJ has a very different correspondence between angular momentum configuration 
and polar anisotropy parameter: for example, a \ChicOne state with $J_z = 0$ or $J_z = \pm 1$ 
leads to $\lth = +1$ and $-1/3$, respectively. 
All polarizations are considered and defined in the centre-of-mass helicity frame.

The polarizations are parametrized as functions of \pTovM by considering for each directly produced state 
its \emph{total} and \emph{longitudinal} cross sections, 
both parametrized as described by Eq.~\ref{eq:powerLaw}.
The ratio of longitudinal to total cross section, the longitudinal fraction, 
is calculated from the polar anisotropy parameter for the two-body decay distribution of the directly produced state,
$\lth^{\mathrm{dir}}(\xi)$, as
$f_{\mathrm{long}}(\xi) = \left[ 1 - \lth^{\mathrm{dir}}(\xi) \right] / \left[ 3 + \lth^{\mathrm{dir}}(\xi) \right]$.
With no further input or prior information than the charmonium data themselves, 
a complete parametrization of cross sections and polarizations of the four considered charmonium states 
(\jpsi, \ChicOne, \ChicTwo and \psip) would require eight $\beta$ parameters, 
describing the \pTovM dependences of the four total direct-production cross sections 
and those of the corresponding longitudinal cross sections.
The remaining eight ``parameters of interest" are the normalizations of the four direct-production cross sections 
and the four corresponding polar anisotropy parameters,
all conventionally considered at the reference point $(\pTovM)^{*} = 5$.

However, not all of the shape parameters are varied independently in the fit. 
Some of them are treated as common to several states and/or polarized subprocesses, 
on the basis of data-driven considerations or basic physics considerations.
The picture of mid-rapidity differential cross sections and polarizations shows, in fact, a characteristic simplicity.
As discussed in Refs.~\cite{bib:Faccioli:PLB773,bib:Faccioli:EPJC78p118}, 
the production cross sections of the ${^3{\rm S}_1}$ and ${^3{\rm P}_J}$ quarkonium states 
measured by ATLAS and CMS, at both 7 and 13\,TeV, 
follow remarkably uniform patterns as a function of \pTovM. 
Such scaling is expected, from dimensional analysis considerations~\cite{bib:Faccioli:EPJC78p118}, 
if the fundamental production processes are identical, in quality and relative contributions, for all states, 
which is a conceivable scenario when we only consider states of identical quantum numbers.
However, an interesting, albeit unexpected, aspect of such ``universal'' picture of mid-rapidity production 
is that there are currently no indications of a difference between the \pTovM distribution shapes 
of the P-wave states and those of the S-wave states~\cite{bib:Faccioli:EPJC78p268}.
Moreover, the measured charmonium and bottomonium decay distributions 
indicate similar polarizations (\lth in the helicity frame) for all vector states, 
independently of their different feed-down contributions from \Chic and \Chib states.
Finally, all polarizations are perfectly compatible with being independent of \pTovM.

These indications of uniform kinematic behaviours are particularly significant 
when we consider the large mass variation between the charmonium and bottomonium families. 
Moreover, the uniformity of such scaling patterns has been verified 
at both 7 and 13\,TeV~\cite{bib:Faccioli:EPJC78p118}.
It is, therefore, reasonable to adopt such quarkonium-wide observations as 
simplifying assumptions for the parametrization of our charmonia-only fit.

We start by imposing that the \emph{directly-produced} vector mesons \jpsi and \psip 
have identical \pTovM-dependent kinematic patterns, while keeping, obviously, 
two independent yield normalizations at $(\pTovM)^{*}$.
This assumption, 
suggested and supported by the observed universality of the \pTovM-differential cross sections and polarizations, 
is also adopted by construction in theoretical models based on the factorization hypothesis (such as NRQCD), 
where the kinematics-dependent short distance cross section terms 
depend only on the heavy quark mass but not on the final bound state~\cite{bib:NRQCD}.
Therefore, in the fit, a single parameter, $\beta_{\mathrm{total}}^{\psi}$, 
describes the \pTovM dependences of the total \jpsi and \psip direct-production cross sections, 
while another one, $\beta_{\mathrm{long}}^{\psi}$, 
represents the shapes of the two corresponding longitudinal cross sections. 
The assumption that both states have the same \emph{direct-production}
polarization is reflected in the choice of one common free parameter for the
polar anisotropy parameter at the reference point, $\lth^{\psi,\mathrm{dir}}(\xi^{*})$.
These constraints ensure that any difference observed 
between the \jpsi and \psip \pTovM distributions and polarizations 
is attributed to the \ChicOne and \ChicTwo feed-down contributions: 
\jpsi and \psip measurements become, therefore, indirect constraints 
on the \ChicOne and \ChicTwo cross sections and polarizations.

Given the relatively large uncertainties affecting some of the data sets used in the analysis,
most notably the \ChicJ cross sections and the \psip polarization, 
it seems judicious to refrain from having many free parameters in the fit model, 
at least in the ``default" analysis.
Therefore, we further require that the production cross sections of all \emph{four} charmonium states 
follow explicitly the \pTovM universality suggested by the ensemble of mid-rapidity quarkonium data: 
one common parameter describes the asymptotic power-law behaviour of their total cross sections,
$\beta_{\mathrm{total}}^{\ChicOne} = 
\beta_{\mathrm{total}}^{\ChicTwo} = 
\beta_{\mathrm{total}}^{\psi}$.
In order not to limit the range of possible physical outcomes, 
no further constraint is imposed on the polarizations:
three independent parameters represent the polar anisotropies of the directly produced states
at the reference $(\pTovM)^{*}$:
$\lth^{\psi,\mathrm{dir}}(\xi^{*})$,
$\lth^{\ChicOne,\mathrm{dir}}(\xi^{*})$, and
$\lth^{\ChicTwo,\mathrm{dir}}(\xi^{*})$.
Furthermore, 
three independent $\beta$ exponents characterize the shapes of the longitudinal cross sections 
of the directly produced states: 
$\beta_{\mathrm{long}}^{\psi}$ (common to the \jpsi and \psip states), 
$\beta_{\mathrm{long}}^{\ChicOne}$ and
$\beta_{\mathrm{long}}^{\ChicTwo}$.
The set of ``parameters of interest" is completed by the normalizations 
of the four (total) direct production cross sections,
defined in the fit model as the $\dd \sigma/\dd\pt$ values (in nb/GeV) at $(\pTovM)^{*}$.
It is worth noting that the measured cross sections and cross section ratios have been published 
in the form of products of the production cross sections and 
branching fractions into the detected decay channel,
while our analysis always considers the pure production cross sections,
obtained dividing the measured values by the relevant branching fractions,
taken from Ref.~\cite{bib:PDG}.

The baseline fit model we have just described has four free $\beta$ exponents 
and, hence, will be referred to in the remaining of this article by the ``$4\beta$" label.
We have also repeated the analysis in two alternative fit configurations, 
analogously labelled as the ``$6\beta$" and ``$1\beta$" models.
In the more unconstrained $6\beta$ scenario, the \ChicOne and \ChicTwo total cross sections 
are free to have power-law \pt trends different from each other and from that of the 
\jpsi and \psip mesons, so that the results will provide a test of the importance of the 
\pTovM-universality we have assumed in the baseline model.
The more constrained $1\beta$ variant imposes a common power-law exponent 
on all total \emph{and longitudinal} cross sections, 
so that the four charmonia are assumed to have identical (fully universal) \pTovM shapes,
not only in the differential cross sections but also in the polarizations.
In other words, in this scenario 
only the \emph{magnitudes} of the cross sections and polarizations can be different 
among the four (directly-produced) states,
being therefore an effective way to directly obtain \pTovM-averaged values 
of the polarizations, of the yield ratios, and of the feed-down fractions.
The number of free power-law exponents is the only difference between 
the $4\beta$, $6\beta$ and $1\beta$ variants, 
the three longitudinal fractions and four direct cross section normalizations at $(\pTovM)^{*}$ 
remaining free parameters in all options.

While the \jpsi and \psip polarization measurements impose, as discussed, 
indirect constraints on the \ChicOne and \ChicTwo polarizations (mainly on their sum), 
direct constraints are provided (mainly on their difference) 
by the three \ChicTwo over \ChicOne yield ratios versus \abscosth,
measured by CMS in three ranges of \jpsi transverse momentum. 
Those data points are parametrized with the expression
\begin{equation}
\label{eq:ChicCosthetaRatio}
R_i \; \frac{ 1 + \lthTwo(\xi_i) \cos^2 \vartheta } { 1 + \lthOne(\xi_i) \cos^2 \vartheta } \, ,
\end{equation}
where $\xi_i$ ($i=1,2,3$) are the three average-\pTovM values of the measurement, 
$R_i$ are the three ratio normalizations, treated as independent fit parameters, 
and $\lth^{\ChicJ}(\xi_i)$ ($J=1,2$) are the polar anisotropy parameters 
of the \jpsi from \ChicOne or \ChicTwo, 
calculated at the three \pTovM values using the parametrized longitudinal and total 
\ChicOne or \ChicTwo cross sections, defined above, 
also including the small \psip feed-down contribution.

Correlations between the data points are taken into account by defining a number of nuisance parameters. 
First, independently for ATLAS and CMS, all the cross sections are scaled by a global factor that, 
while being a free parameter in the fit, is constrained by a Gaussian function of mean unity 
and width equal to the relative uncertainty of the integrated luminosity, reported in the experimental publications.
In other words, the fit quality incurs a penalty reflecting the difference between the best-fit scale factor and unity,
normalized by the uncertainty. 
By equally scaling all the cross sections of a given experiment, these two nuisance parameters induce a correlation 
between the \psip, \ChicOne and \ChicTwo cross sections measured by ATLAS and
between the \jpsi and \psip cross sections measured by CMS.
Second, also the branching factions needed to convert the measured values to production cross sections are 
analogously scaled by Gaussian-constrained nuisance parameters, 
the Gaussian widths being the relative uncertainties reported in Ref.~\cite{bib:PDG}.
This second set of nuisance parameters induces correlations between the ATLAS and CMS data.
It turns out that all the post-fit nuisance parameters are identical to unity,
except for the one related to the $\psip \to \mu\mu$ branching fraction, 
which deviates from unity by 1\%, 
a negligible departure given that the uncertainty is six times larger.

Another, very important, source of correlations between all data points 
is the dependence of the detection acceptances on the polarization.
For each set of parameter values considered in the fit scan, 
the expected values of the polarizations and cross sections are calculated, 
for all states, as functions of \pt, using the shape-parametrization functions described above.
The expected \lth values can be immediately compared to the measured ones, 
for the determination of the corresponding $\chi^2$ terms, 
while for the calculation of the cross section $\chi^2$ terms we first scale the measured cross sections 
by acceptance-correction factors calculated for the \lth value under consideration. 
These correction factors are computed, for each data point, using the tables published by the experiments 
(for exactly this purpose) for the cross sections of particles produced with fully transverse or fully longitudinal polarizations.

\section{Results of the global fit analysis}

As can be appreciated from the information presented in Table~\ref{tab:fit_quality},
the fit quality is excellent in all three fit variants.
No tension or difference in trends is visible between the 108 data points and the best fit curves, 
as shown in Figs.~\ref{fig:xsections}, \ref{fig:lth} and~\ref{fig:Rchic_vs_costh} for the baseline $4\beta$ case.
The smallness of the $\chi^2$ per degree of freedom, $\chi^2/{\rm ndf} = 40/93$, 
corresponding to an exceptionally (and suspiciously) good fit $\chi^2$ probability,
points to the existence of unaccounted correlations between systematic uncertainties in the data points.
Indeed, it is very likely that a fraction of the systematic uncertainties assigned in the experimental publications
to each of the \pt bins actually reflects an effect that commonly affects a broad region of the 
distribution,
leaving its shape essentially unchanged,
so that the true point-to-point uncorrelated uncertainties are somewhat smaller than those we have used.
In any case, it is certainly informative to compare the $\chi^2/{\rm ndf}$ value of the baseline analysis
with those of the two variants mentioned before: $\chi^2/{\rm ndf} = 39/91$ ($6\beta$) and 43/96 ($1\beta$). 
We see that, given the precision of the presently-available experimental inputs, 
there is no advantage in using the fit model with two more free parameters.
Indeed, according to the Akaike information criterion (AIC)~\cite{bib:AIC},
the likelihood of the $6\beta$ model is much smaller than that of the $4\beta$ model.

\begin{table}[ht]
\centering
\caption{Fit quality information. 
In all three fit variants there are 108 data points and 10 nuisance parameters.}
\label{tab:fit_quality}
\begin{tabular}{lccc}
\hline\noalign{\vglue0.85mm}
 & $6\beta$ & $4\beta$ & $1\beta$ \\ 
\noalign{\vglue0.85mm}\hline\noalign{\vglue0.85mm}
No.\ of free parameters & 27 & 25 & 22 \\
No.\ of degrees of freedom & 91 & 93 & 96 \\
Fit $\chi^{2}$ & 39 & 40 & 43 \\
\noalign{\vglue0.85mm}\hline
\end{tabular}
\end{table}

Interestingly, the more constrained $1\beta$ fit model, 
which imposes a common value to all the six $\beta$ exponents,
provides a description of the data that is essentially as good as that of the baseline option,
despite having three less free parameters.
The slightly worse fit $\chi^2$ is compensated by the extra simplicity of the model,
leading to a large increase in the AIC relative likelihood.
We will refrain from using this observation to highlight the implication that 
all charmonium states are seemingly produced with identical kinematical patterns, 
both in terms of cross sections and in terms of polarizations, so that a rather straightforward 
model is able to faithfully reproduce all the data points considered in our study.
Instead, we simply argue that this remarkable observation should trigger further experimental measurements 
of quarkonium cross sections and polarizations, with significantly improved precision, 
so that the validity of the $1\beta$ model can be scrutinised much more accurately.
Only then we will be able to conclude if this strongly constrained fit is merely
a very effective and economic description of the presently existing data,
providing a reliable computation of \pTovM-averaged results,
or if we are seeing a smoking-gun signature of a fully-universal scenario,
reflecting a deeper symmetry at the core of hadron formation than assumed 
in today's theories of quarkonium production.

\begin{table}[h!]
\centering
\caption{Values of the fitted parameters of interest, in the three considered scenarios.}
\label{tab:best_fit_pars}
\begin{tabular}{lccc}
\hline\noalign{\vglue0.85mm}
 & $6\beta$ & $4\beta$ & $1\beta$ \\ 
\noalign{\vglue0.85mm}\hline\noalign{\vglue0.85mm}
    \multicolumn{4}{c}{Direct-production $\dd \sigma/\dd\pt$\,(nb/GeV), at $\pTovM = 5$}\\
\noalign{\vglue0.85mm}\hline\noalign{\vglue0.85mm}
    \jpsi          & $1.274 \pm 0.059$  & $1.283 \pm 0.059$  & $1.281 \pm 0.058$ \\
    \psip         & $0.230 \pm 0.010$  & $0.232 \pm 0.010$  & $0.233 \pm 0.010$ \\
    \ChicOne  & $1.008 \pm 0.097$  & $0.966 \pm 0.092$  & $1.008 \pm 0.085$ \\
    \ChicTwo  & $0.604 \pm 0.071$  & $0.577 \pm 0.068$  & $0.617 \pm 0.063$ \\
\noalign{\vglue0.85mm}\hline\noalign{\vglue0.85mm}
  \multicolumn{4}{c}{Direct \lth at $\pTovM = 5$}\\
\noalign{\vglue0.85mm}\hline\noalign{\vglue0.85mm}
  $\lth^{\psi,\rm dir}$         & $-0.005 \pm 0.072$ &  $\hspace{\digitwidth} 0.022 \pm 0.062$ &  $\hspace{\digitwidth} 0.040 \pm 0.060$ \\
\noalign{\vglue0.85mm}
  $\lth^{\chi_{c1},\rm dir}$ & $\hspace{\digitwidth} 0.504 \pm 0.303$ &  $\hspace{\digitwidth} 0.371 \pm 0.268$ &  $\hspace{\digitwidth} 0.521 \pm 0.247$ \\
\noalign{\vglue0.85mm}
  $\lth^{\chi_{c2},\rm dir}$ & $-0.402 \pm 0.276$ & $-0.533 \pm 0.250$ & $-0.392 \pm 0.233$ \\
\noalign{\vglue0.85mm}\hline\noalign{\vglue0.85mm}
    \multicolumn{4}{c}{Kinematical-dependence parameters}\\
\noalign{\vglue0.85mm}\hline\noalign{\vglue0.85mm}
    $\gamma$                  & $0.642 \pm 0.176$  & $0.643 \pm 0.157$  & $0.601 \pm 0.148$  \\
\noalign{\vglue0.85mm}
    $\beta_{\mathrm{total}}^{\psi}$        & $3.358 \pm 0.032$  & 
    \multirow{3}{*}{\hspace{-5pt}$\left.\rule[1.5ex]{0pt}{3.5ex}\right\} 3.379 \pm 0.022$}  & 
    \multirow{6}{*}{\hspace{-5pt}$\left.\rule[1.5ex]{0pt}{8.8ex}\right\} 3.385 \pm 0.021$}  \\
\noalign{\vglue0.85mm}
    $\beta_{\mathrm{total}}^{\chi_{c1}}$   & $3.444 \pm 0.122$  &   &   \\
\noalign{\vglue0.85mm}
    $\beta_{\mathrm{total}}^{\chi_{c2}}$   & $3.562 \pm 0.171$  &   &   \\
\noalign{\vglue0.85mm}
    $\beta_{\mathrm{long}}^{\psi}$        & $3.494 \pm 0.142$  & $3.431 \pm 0.097$  &   \\
\noalign{\vglue0.85mm}
    $\beta_{\mathrm{long}}^{\chi_{c1}}$   & $3.178 \pm 0.715$  & $3.530 \pm 0.599$  &   \\
\noalign{\vglue0.85mm}
    $\beta_{\mathrm{long}}^{\chi_{c2}}$   & $3.429 \pm 0.322$  & $3.674 \pm 0.273$  &   \\
\noalign{\vglue0.85mm}\hline
\end{tabular}
\end{table}

\begin{figure}[t!]
\centering
\includegraphics[width=0.98\linewidth]{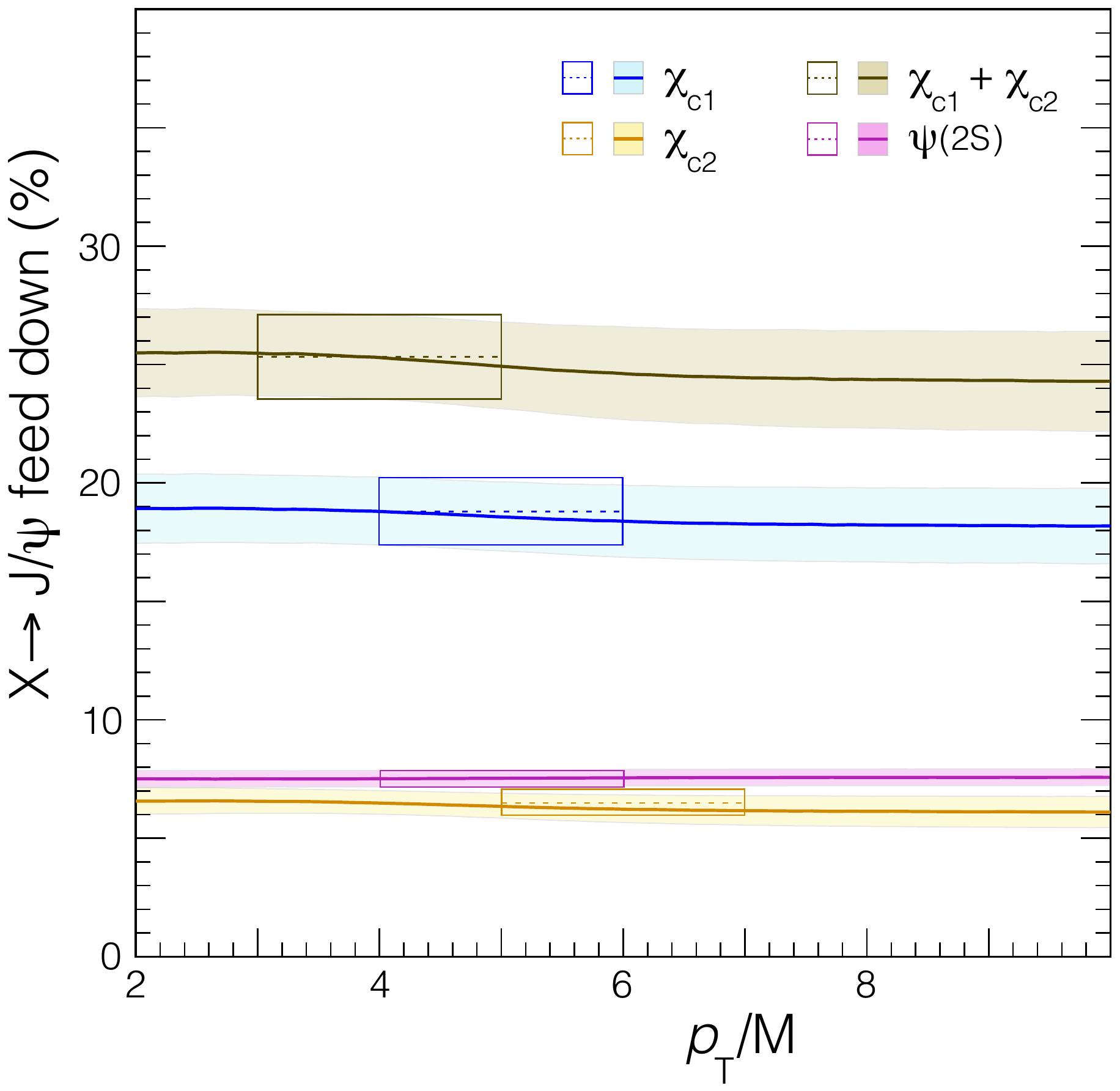}
\caption{The fractions of the total prompt \jpsi production rate due to 
feed-down decays from the \ChicOne, \ChicTwo and \psip mesons,
as a function of \pTovM.}
\label{fig:psi_FD_vs_pTovM}
\end{figure}

The fitted values of all the parameters of interest are collected in Table~\ref{tab:best_fit_pars},
for the three variants: $6\beta$, $4\beta$ and $1\beta$.
The obtained \Chic-to-\jpsi and \psip-to-\jpsi feed-down fractions
are shown in Fig.~\ref{fig:psi_FD_vs_pTovM}.
The results of the baseline ($4\beta$) fit are represented by the \pTovM-dependent 
central values (solid lines), enveloped by filled bands of widths equal to the 
68.3\% confidence level uncertainties, 
obtained by integrating the multivariate normal distribution representing the 
joint probability distribution of all parameters over the physical domains 
of all the variables not shown in the figure.
The results of the $1\beta$ fit option, independent of \pTovM by construction, 
are also shown in the figure, 
as dashed lines (central values) surrounded by empty rectangles (uncertainties);
the corresponding numerical values are collected in Table~\ref{tab:FD_fractions}.

\begin{table}[t]
\centering
\caption{\pTovM-averaged values of the feed-down fractions, 
as determined in the $1\beta$ global-fit analysis. 
Virtually identical values are obtained in the $4\beta$ fit option, for $\pTovM = 5$.
The derived direct \jpsi fraction is $(67.2 \pm 1.9)$\%.}
\label{tab:FD_fractions}
\begin{tabular}{lc}
\hline\noalign{\vglue0.85mm}
\multicolumn{2}{c}{Feed-down fractions (\%)} \\ 
\noalign{\vglue0.85mm}\hline\noalign{\vglue0.85mm}
$\ChicOne \to \jpsi$ & $18.8 \pm 1.4$ \\[2.5pt]
$\ChicTwo \to \jpsi$ & $\hspace{\digitwidth} 6.5 \pm 0.5$ \\[2.5pt]
$\ChicOne + \ChicTwo \to \jpsi$  & $25.3 \pm 1.8$ \\[2.5pt]
$\psip \to \jpsi$  & $\hspace{\digitwidth} 7.5 \pm 0.3$ \\[2.5pt]
$\psip \to \ChicOne$ & $\hspace{\digitwidth} 2.2 \pm 0.2$ \\[2.5pt]
$\psip \to \ChicTwo$ & $\hspace{\digitwidth} 3.4 \pm 0.3$ \\
\noalign{\vglue0.85mm}\hline
\end{tabular}
\end{table}

\begin{figure*}[t!]
\centering
\includegraphics[width=0.85\linewidth]{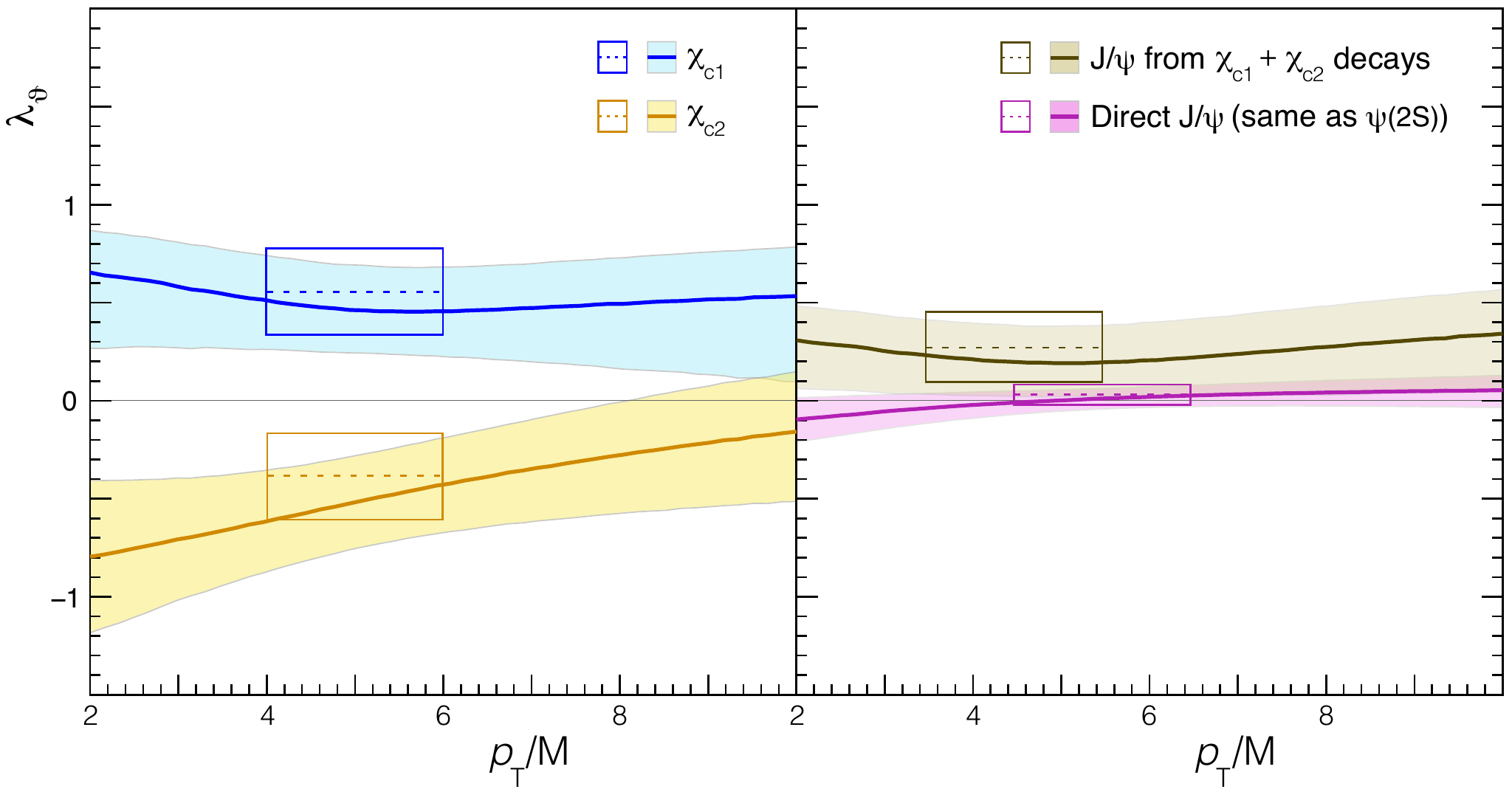}
\caption{The polarization parameter \lth of the \ChicOne and \ChicTwo mesons (left),
as well as of the \jpsi mesons directly produced (same as of the \psip)
and produced in \ChicOne plus \ChicTwo decays (right),
as a function of \pTovM.}
\label{fig:lth_chic_vs_pTovM}
\end{figure*}

The corresponding results for the polarizations (\lth in the helicity frame) 
are presented in Fig.~\ref{fig:lth_chic_vs_pTovM}, 
in the left panel for the \ChicOne and \ChicTwo mesons and 
in the right panel for the directly-produced \jpsi mesons.
The right panel also shows the polarization of the \jpsi mesons
produced in decays of both \Chic mesons, 
an observable determined with a better precision than each of the individual 
(anti-)correlated polarizations shown on the left panel.
As in Fig.~\ref{fig:psi_FD_vs_pTovM}, 
the dashed lines and empty rectangles represent the \pTovM-independent results
obtained in the $1\beta$ fit,
effectively representing averages over \pTovM of the $4\beta$ fit results,
shown as solid lines and filled bands.
The respective numerical values are collected in Table~\ref{tab:lth_values},
which also shows the derived polarization of promptly produced \jpsi\ mesons,
naturally intermediate between the values of the directly produced mesons 
and of those emitted in the \ChicJ decays.

\begin{table}[ht]
\centering
\caption{Polarizations determined in the $4\beta$ fit variant (for $\pTovM = 5$) and
in the $1\beta$ option (averaged over \pTovM).}
\label{tab:lth_values}
\begin{tabular}{lcc}
\hline\noalign{\vglue0.85mm}
\multicolumn{3}{c}{$\lambda_{\vartheta}$} \\ 
\noalign{\vglue0.85mm}\hline\noalign{\vglue0.85mm}
 & $4\beta$ & $1\beta$ \\ 
\noalign{\vglue0.85mm}\hline\noalign{\vglue0.85mm}
\ChicOne  & $\hspace{\digitwidth} 0.46 \pm 0.23$ & $\hspace{\digitwidth} 0.55 \pm 0.23$ \\[5pt]
\ChicTwo  & $-0.52 \pm 0.24$ & $-0.39 \pm 0.22$ \\[5pt]
\jpsi from $\ChicOne + \ChicTwo$ & $\hspace{\digitwidth} 0.19 \pm 0.18$ & $\hspace{\digitwidth} 0.27 \pm 0.19$ \\[5pt]
Direct \jpsi = \psip  & $0.022 \pm 0.062$ & $\hspace{\digitwidth} 0.040 \pm 0.060$ \\[5pt]
Prompt \jpsi   & $\hspace{\digitwidth} 0.048 \pm 0.037$ & $\hspace{\digitwidth} 0.087 \pm 0.024$ \\
\noalign{\vglue0.85mm}\hline
\end{tabular}
\end{table}

\begin{figure*}[t!]
\centering
\includegraphics[width=0.9\linewidth]{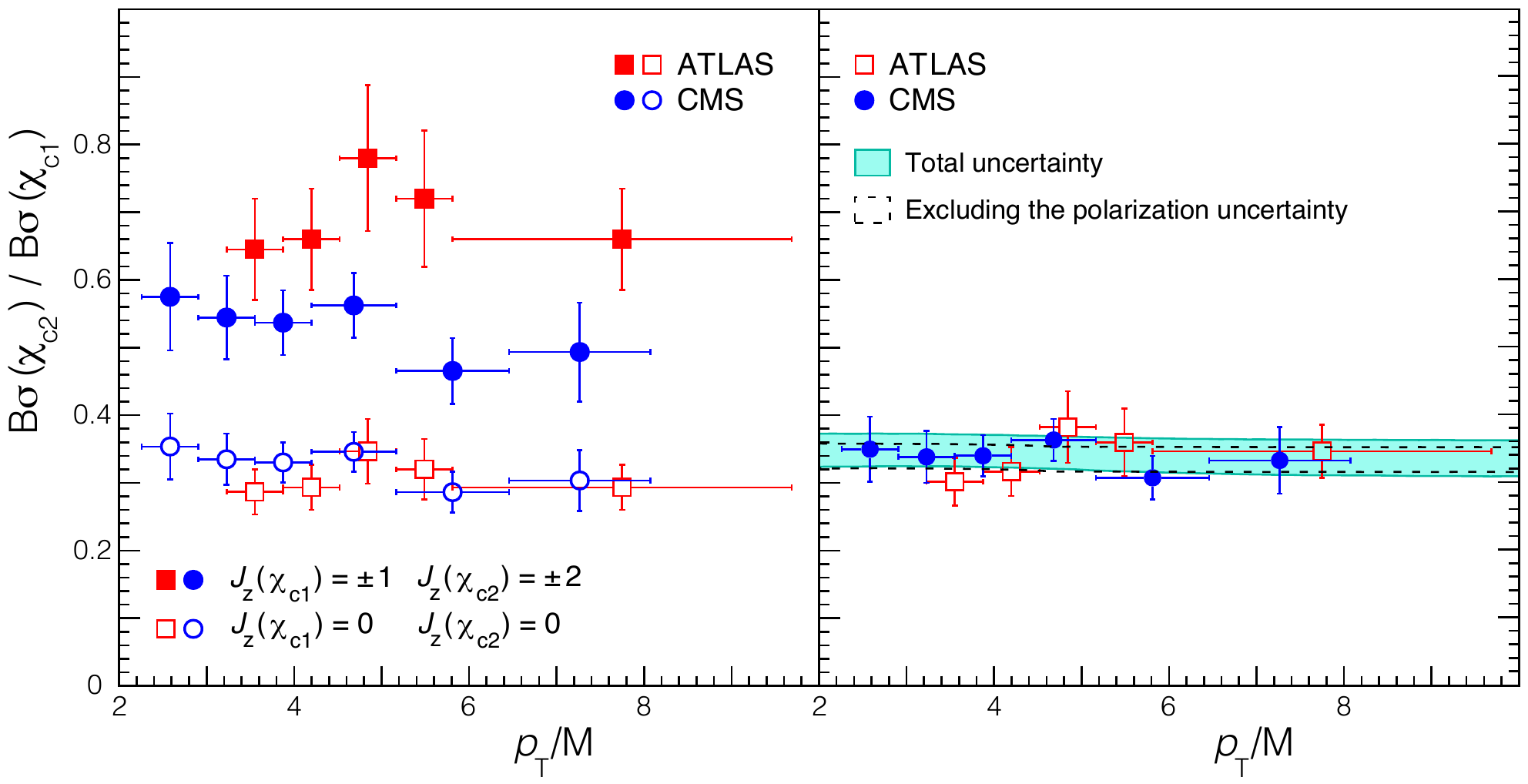}
\caption{Ratio between the \ChicTwo and \ChicOne cross sections, 
times the corresponding $\ChicJ \to \jpsi \, \gamma$ branching fractions, 
as a function of \pTovM, for two extreme polarization scenarios (left)
and for the polarizations determined in our global-fit analysis (right).}
\label{fig:Rchic_vs_pTovM}
\vglue5pt
\centering
\includegraphics[width=0.48\linewidth]{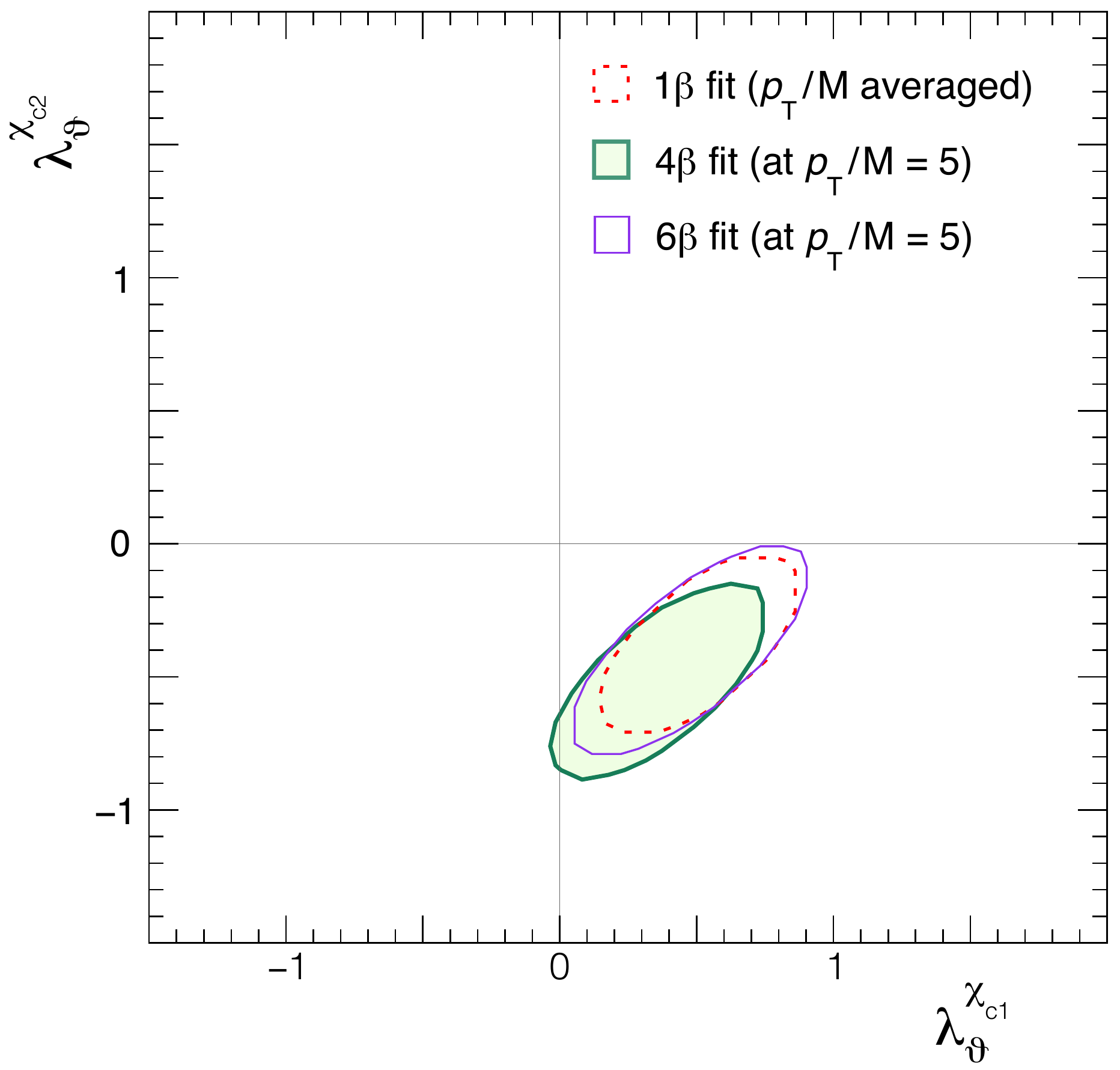}
\hfil
\includegraphics[width=0.48\linewidth]{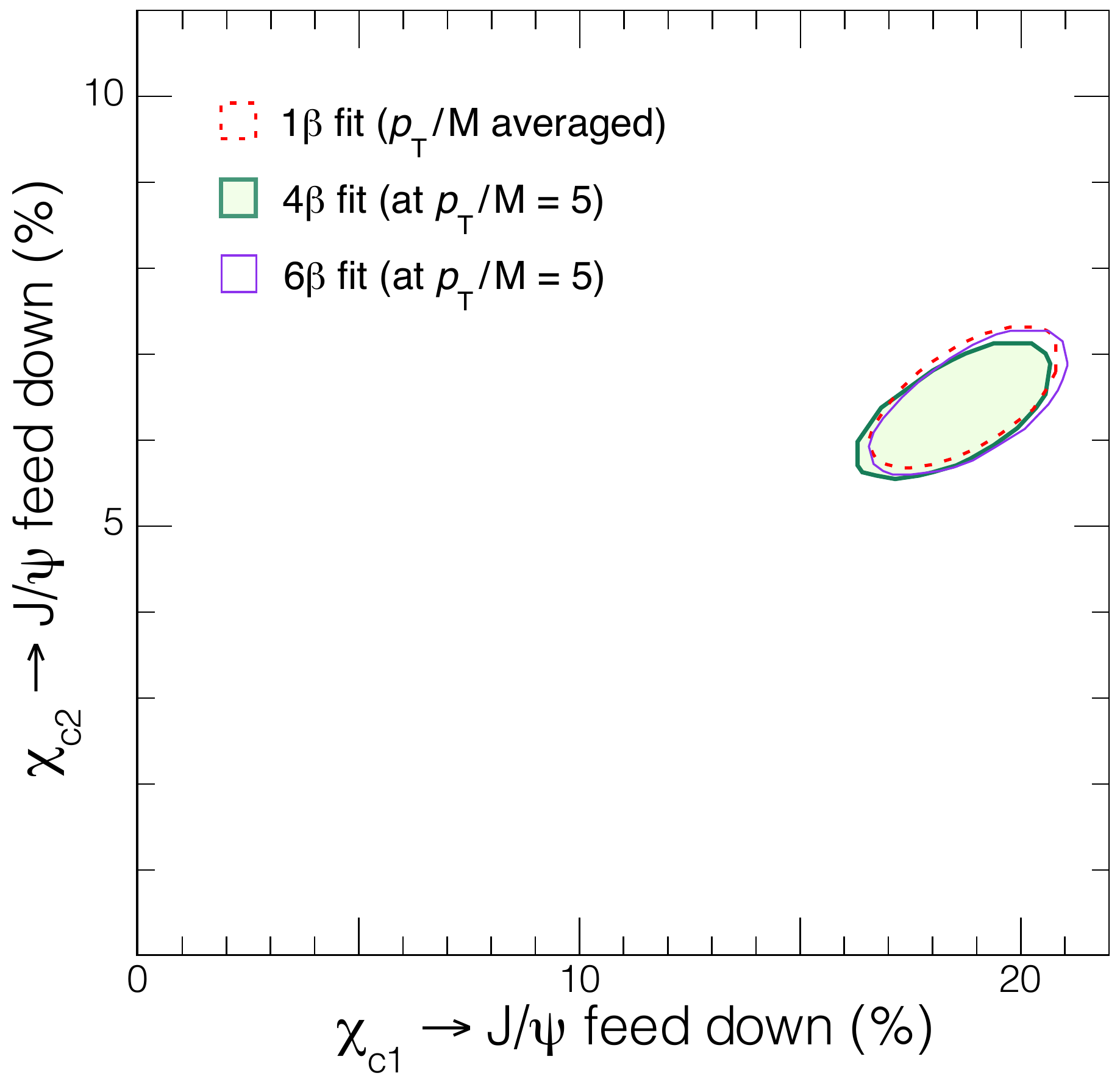}
\caption{Two-dimensional distributions showing the correlations 
between the \lthTwo and \lthOne polarizations (left) and 
between the \ChicTwo and \ChicOne feed-down contributions to the prompt \jpsi production rate (right).
The results of the three fit variants are represented by the 
solid ($4\beta$), dashed ($1\beta$) and dotted-dashed ($6\beta$) lines.}
\label{fig:2D_chic2_vs_lth_chic1_lth_and_FD}
\end{figure*}

Our global-fit analysis provides significant improvements in the determination of several interesting observables. 
To start with, individual (purely data-driven) values of the \ChicOne and \ChicTwo polarizations are extracted, 
as reported in Table~\ref{tab:lth_values}.
Equally important, the feed-down fractions (shown in Table~\ref{tab:FD_fractions})
are determined with a rather good precision, of around 10\%, 
even for the small fractions of \ChicOne and \ChicTwo production yields due to radiative decays of \psip mesons.
Finally, the ratio between the \ChicTwo and the \ChicOne cross sections
(times the corresponding $\ChicJ \to \jpsi \, \gamma$ branching fractions),
becomes much more precisely determined:
$B\sigma(\ChicTwo) / B\sigma(\ChicOne) = 0.343 \pm 0.024$.
Figure~\ref{fig:Rchic_vs_pTovM} provides a graphical illustration of the
improvement reached in the precision of the \ChicTwo to \ChicOne cross section ratio.
The left panel shows the very strong dependence of the original ATLAS and CMS measurements  
(and of the level of their mutual compatibility) on the unknown \ChicJ polarizations.
The \ChicJ polarization constraints contributing (directly and indirectly) to our global fit
strongly reduce the uncertainty associated with the polarization dependence of the acceptance correction, 
leading to the rather well aligned points shown on the right panel,
where the acceptance corrections reflect the best-fit polarization scenario 
(curiously, very close to the $J_z = 0$ extreme).
The filled band represents the final result and
the difference between the widths of the filled and dashed bands reflects the residual polarization uncertainty,
a rather small effect in comparison with the impact seen in the left panel.

Among the remaining physical results, 
particularly interesting is the polarization of the \psip and of the directly produced \jpsi,
shown by the pink band in Fig.~\ref{fig:lth_chic_vs_pTovM},
constrained to be ``zero" with a previously unseen precision and no signs of momentum dependence. 
This is a unique result for a vector state; both Drell--Yan 
dileptons~\cite{Lam:1978pu,bib:Faccioli:PRLgeneralizeLT,Guanziroli:1987rp,Conway:1989fs,Zhu:2006gx,Zhu:2008sj}
and vector boson~\cite{Mirkes:1994eb,Aaltonen:2011nr,Khachatryan:2015paa,Abbott:2000aj,
Acosta:2003jm,Acosta:2005dn,Chatrchyan:2011ig,ATLAS:2012au}
polarizations are known to be significantly non-zero and momentum dependent, 
as are those of low-\pt quarkonia~\cite{Abt:2009nu,Brown:2000bz}.
In fact, there are only two ways to obtain a vector particle in an angular momentum state 
having zero observable polarization. 
One is to prepare a mixture of two (or more) very different, strongly polarized states: 
as demonstrated in Ref.~\cite{bib:Faccioli:PRLgeneralizeLT}, 
for a \emph{single}, individually produced angular momentum state 
there is always a polarization axis with respect to which $\lth = +1$. 
The \emph{exact} compensation of two strongly polarized production processes, 
leaving no margin for a residual momentum-dependent deviation of \lth from zero, 
would be an astonishing coincidence; 
in fact, it could only be attributed to the existence of unknown symmetries governing charmonium production, 
at least in the mid-rapidity limit~\cite{bib:Faccioli:NRQCD2019}. 
The other possibility is that the \jpsi originates from the decay of a $J = 0$ state, 
as expected to happen in the production from the $^1{\rm S}_0$ colour octet term in NRQCD 
(and in the feed-down from \ChicZero). 
In this possible subprocess, 
while the polarization continues to be naturally fully transverse along the direction of the recoil gluon (or photon), 
it undergoes a complete rotational smearing when seen in the experimental polarization frame, 
whose $z$ axis is fully decorrelated from such natural direction when the mother-daughter mass difference is small. 
While the production via $^1{\rm S}_0$ octet is foreseen, 
it is not naturally predicted to be the only, dominating mechanism. 
In either case, a precise confirmation of a \pt-independent unpolarized scenario has strong 
and rather remarkable physical implications. 

\begin{figure*}[t!]
\centering
\includegraphics[angle=270,width=\linewidth]{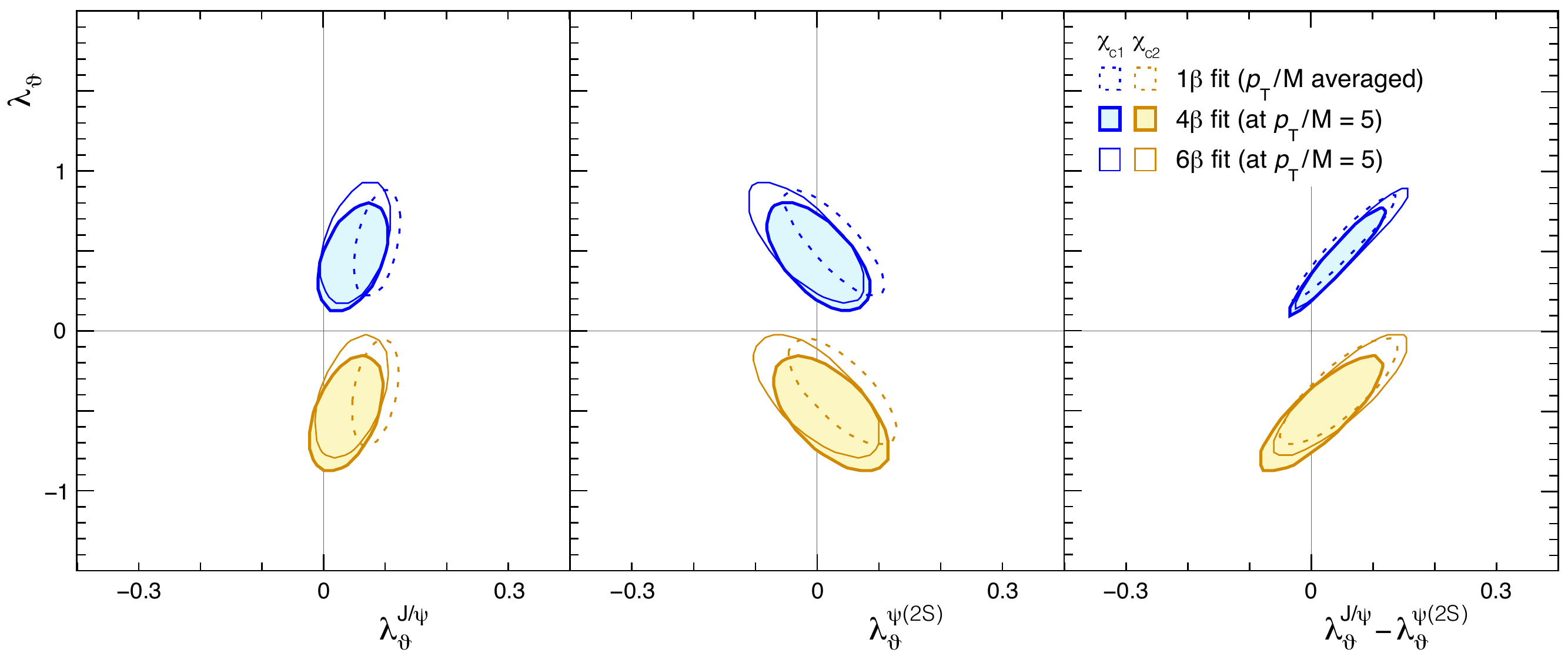}
\caption{Two-dimensional distributions showing the correlations between 
the \lth polarization parameters of the \ChicOne or \ChicTwo mesons
and the corresponding values for the \jpsi (left), the \psip (middle) 
and the difference between the two (right).
The results of the three fit variants are represented by the 
solid ($4\beta$), dashed ($1\beta$) and dotted-dashed ($6\beta$) lines.}
\label{fig:lth_chic12_vs_lth_psi}
\end{figure*}

\begin{figure}[ht]
\centering
\includegraphics[width=0.98\linewidth]{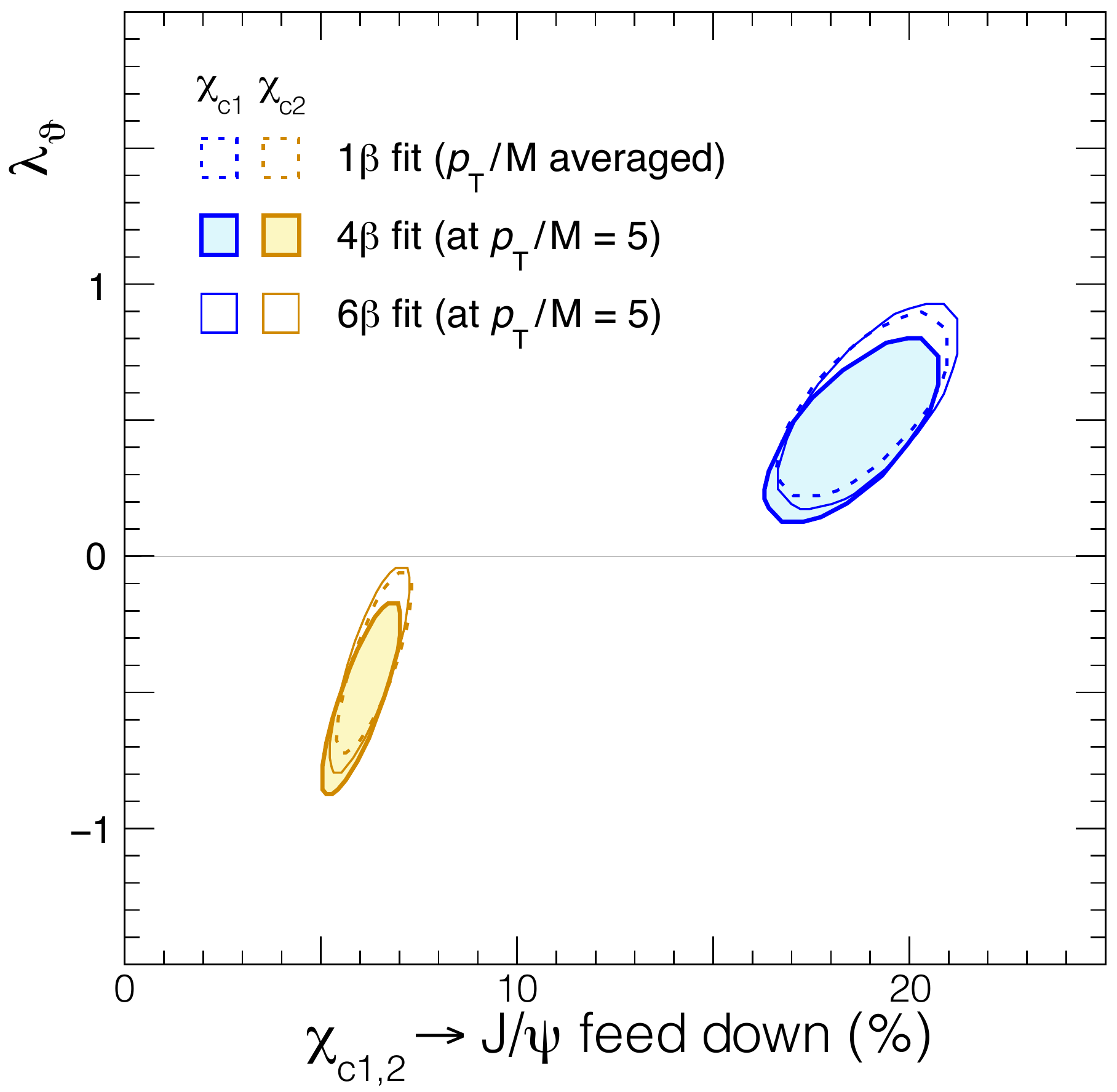}
\caption{Two-dimensional distributions showing the correlation between the \lthOne (\lthTwo) 
observable and the \ChicOne (\ChicTwo) feed-down contribution to the prompt \jpsi production rate.
The results of the three fit variants are represented by the 
solid ($4\beta$), dashed ($1\beta$) and dotted-dashed ($6\beta$) lines.}
\label{fig:lth_vs_FD}
\end{figure}

The variations represented by the bands in Figs.~\ref{fig:psi_FD_vs_pTovM} and~\ref{fig:lth_chic_vs_pTovM}
are generally correlated. 
Figure~\ref{fig:2D_chic2_vs_lth_chic1_lth_and_FD}
shows the correlations between the \ChicOne and \ChicTwo polarizations (left)
and between the \ChicOne and \ChicTwo feed-down contributions to \jpsi production (right),
for the three fit variants.
Particularly interesting are the correlations shown in Fig.~\ref{fig:lth_chic12_vs_lth_psi}, 
where we can see that a significantly improved knowledge of the \ChicJ polarizations 
will derive from new, precise measurements of the \psip polarization and, above all, 
of the difference between the \jpsi and \psip polarizations. 
This latter measurement can be performed, for example, 
by determining the ratio between the \jpsi and \psip angular distributions in a given \pTovM interval, 
with the cancellation of a large part of the important systematic uncertainties 
related to acceptance and efficiency descriptions. 
It would, therefore, represent a particularly clean constraint on the sum of the \ChicOne and \ChicTwo polarizations.
Finally, Fig~\ref{fig:lth_vs_FD} shows that the \ChicOne and \ChicTwo polarizations are also correlated with 
the corresponding feed-down fractions to \jpsi production, 
so that precise measurements of those feed-down fractions will also reduce 
the \lthOne and \lthTwo uncertainties.

\section{Summary}

We have performed a global study of charmonium production at mid-rapidity and LHC energies, 
with the aim to improve the current knowledge of \ChicJ polarizations and feed-down fractions to \jpsi production,
and to extract the kinematic properties of the directly produced \jpsi.
The analysis is fully data driven, not relying on any theoretical inputs.
It uses LHC data on \jpsi, \psip, \ChicOne and \ChicTwo differential cross sections and decay angular distributions,
measured by ATLAS and CMS at mid-rapidity, in pp collisions at $\sqrt{s} = 7$\,TeV.

A first result of the analysis is that all polarizations and cross section ratios 
are found to be perfectly compatible with being \pTovM-independent.
When the \pTovM scaling of the cross sections --- 
whose evidence is further strengthened by the particularly significant comparison 
between charmonium and bottomonium data at 7 and 13\,TeV --- 
is imposed as direct constraint in the fit, 
the feed-down fractions, polarizations and cross section ratios are determined with good precision.
We note that the \jpsi feed-down fractions are perfectly compatible with values obtained 
in a global analysis of low-\pt data from fixed-target experiments~\cite{bib:Faccioli:FD2009},
an observation that confirms the independence of such ratios on \pTovM and collision energy.

\begin{figure}[t]
\centering
\includegraphics[width=0.98\linewidth]{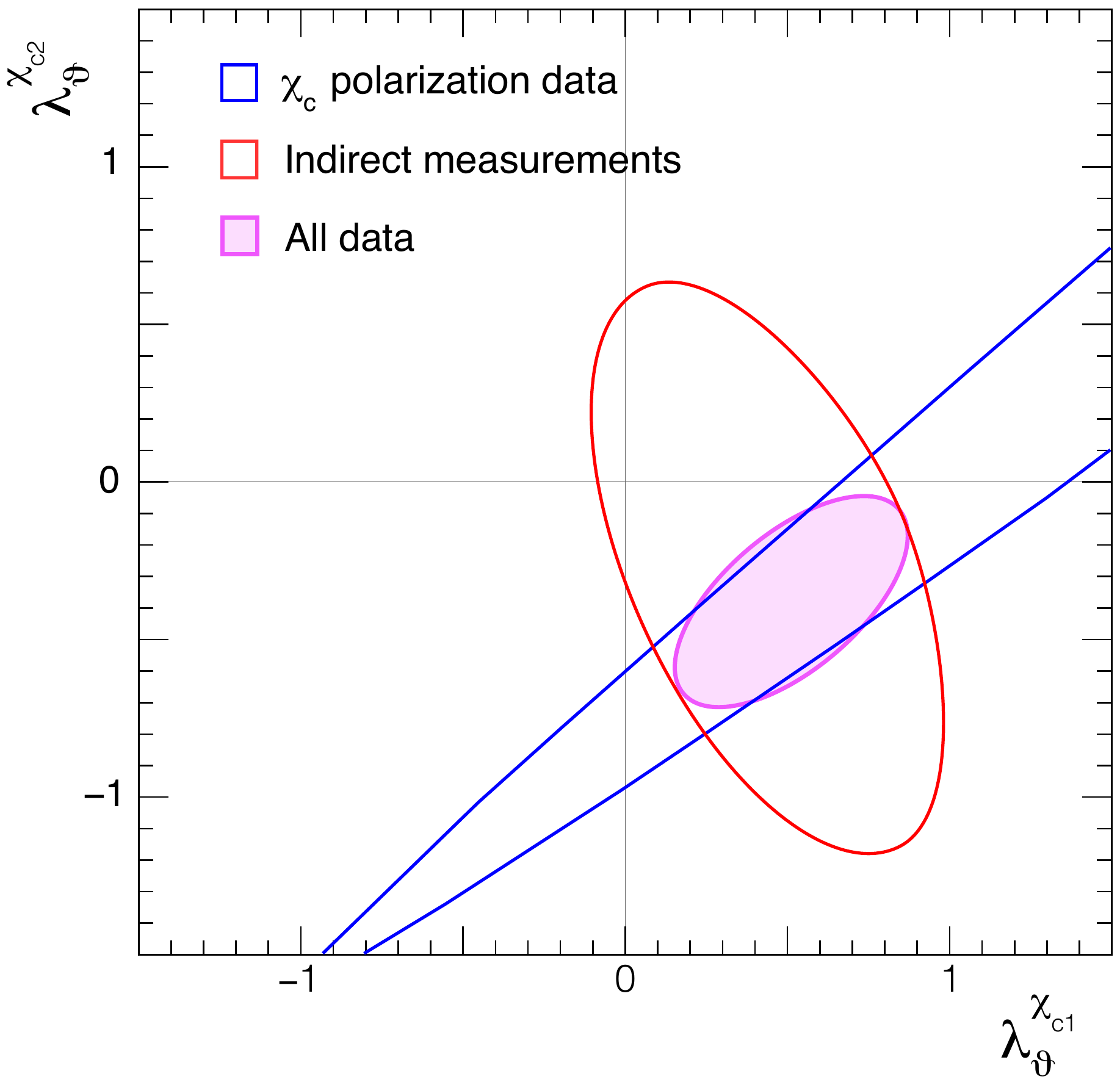}
\caption{Two-dimensional distributions showing the correlation 
between the \lthTwo and \lthOne polarizations, in the $1\beta$ fit variant,
when using three sets of measurements:
only the direct \ChicJ polarization data~\cite{bib:BPH-13-001} (blue), 
which essentially constrains the $\lthTwo - \lthOne$ difference,
all the other data (red), mostly constraining the $\lthOne + \lthTwo$ sum,
and the result of the global fit presented in this paper (pink).}
\label{fig:2D_3cases}
\end{figure}

While the \ChicOne and \ChicTwo individual polarizations remain the least well known observables, 
the global-fit of all available data provides a first determination of their individual values.
The significant improvement in our knowledge can be seen in Fig.~\ref{fig:2D_3cases},
where the result of the global fit (in the $1\beta$ fit variant) reported in this paper (pink contour) 
is compared to the almost orthogonal results obtained with two complementary subsets of constraints: 
the direct \ChicJ polarization data shown in Fig.~\ref{fig:Rchic_vs_costh} (blue line, from Ref.~\cite{bib:BPH-13-001})
and all the indirect experimental information (red line).

Further improvements do not need to come from future \ChicJ polarization measurements,
which are notoriously challenging; 
precise data on the \jpsi and \psip polarizations, as well as on the \ChicJ feed-down fractions, 
can also lead to better determinations of the \ChicJ polarizations.
In particular, a significant improvement can be obtained through the measurement
of the \emph{difference} between the prompt \jpsi and \psip polarizations, 
potentially very precise given the cancellation of most systematic uncertainties.
New measurements of the \psip polarization, especially towards higher \pt, are also a top priority.
In fact, the polarization of \emph{directly produced} \jpsi and \psip states,
accessible for the first time as a result of our global fit analysis,
is found to be very small ($\lthpsi = 0.04 \pm 0.06$) and \pt-independent. 
It is important to clarify if this fine-tuned balance between transverse and longitudinal yields 
is only attained within the relatively narrow \pt window covered by the presently available data, 
in which case it can be seen as a mere coincidence, 
or remains unbroken up to higher \pt values,
in which case it can be seen as a clear sign of a highly peculiar underlying production mechanism, 
probably involving a not yet understood symmetry.

\bigskip
Acknowledgement: 
The work reported in this paper has been performed using an extended version of the 
analysis framework previously used to obtain the results reported in Ref.~\cite{bib:Faccioli:EPJC78p268}, 
a study made in collaboration with M.\ Ara\'ujo, J.\ Seixas, I. Kr{\"a}tschmer and V. Kn{\"u}nz.

\bibliographystyle{cl_unsrt}
\bibliography{references}{}

\end{document}